\documentclass[12pt,a4paper]{article}
\usepackage{graphicx}
\usepackage{amsmath}
\usepackage{amsfonts}
\usepackage{amssymb}
\usepackage{subcaption}
\usepackage{mathptmx}

\newcommand{\ove}{\overline }
\newcommand{\eps}{\varepsilon}

\begin{document}

\begin{center}
{\bfseries Supernova Neutrino Burst Monitor at the Baksan Underground Scintillation Telescope}
\vspace{5mm}

{Yu.F.~Novoseltsev$^{1,}$\footnote{email:\ novoseltsev@inr.ru}, M.M.~Boliev$^1$, I.M.~Dzaparova$^{1,2}$, M.M.~Kochkarov$^1$, A.N.~Kurenya$^1$, R.V.~Novoseltseva$^1$, V.B.~Petkov$^{1,2}$, P.S.~Striganov$^1$,  A.F.~Yanin$^1$}
\vspace{5mm}

  {\small {$^1$ Institute for Nuclear Research of RAS, 117312 Moscow, Russia}\\
         {$^2$ Institute of Astronomy of RAS, 119017 Moscow, Russia}}

\end{center}

%\address{Institute for Nuclear Research of the Russian Academy of Sciences, 60th %October Anniversary Prospect, 7a, 117312 Moscow, Russia}

\begin{abstract}
The experiment on recording neutrino bursts operates since the mid-1980. As the target, we use two parts of the facility with the total mass of 240 tons. The current status of the experiment and some results related to the investigation of background events and the stability of facility operation are presented. Over the period of June 30, 1980 to December 31, 2018, the actual observational time is 33.02 years. No candidate for the stellar core collapse has been detected during the observation period. An upper bound of the mean frequency of core collapse supernovae in our Galaxy is  0.070 year$^{-1}$ (90\% CL).
\end{abstract}

\section{Introduction}
Recording the supernova SN 1987A has made a considerable impact on both theoretical investigation of the SN phenomenon and experimental facilities development. Core collapse supernovae are among the most powerful sources of neutrinos in the
Universe.

The detection of neutrinos from the SN1987A experimentally proved the critical role of neutrinos in the explosion of massive stars,  as it was suggested more than 50 years ago \cite{Gamov,Zeld,Colg}.

Due to their high penetration power, neutrinos deliver information on physical conditions in the core of the star during the gravitational collapse. SN1987A
has become the nearest supernova in the past several hundred years, which allowed the SN formation process to be observed in unprecedented detail beginning
with the earliest time of radiation. It was the first time that a possibility arose for comparing the main parameters of the existing theory -- total radiated energy,
neutrino temperature, and neutrino burst duration -- with the experimentally measured values \cite{Lore,Pagl}.

The SN1987A event has demonstrated significant deviations from spherical
symmetry. It means the SN phenomenon is substantially multidimensional process. In recent years great progress has been achieved in two-dimensional (2D) and three-dimensional (3D) computer simulations of an SN explosion. 3D simulations of the evolution of massive stars at the final stage of their life (SN progenitors)
have revealed very important role of non-radial effects. However, further analysis
would be mandatory when high-resolution 3D-simulations will become available.

Since light (and electromagnetic radiation in general) can be partially or completely absorbed by dust in the galactic plane, the most appropriate tool for finding supernovae with core collapse are large neutrino detectors. In the past decades (since 1980), the search for neutrino bursts was carried out with such detectors as the Baksan Scintillation Telescope \cite{Alek3,Nov1}, Super-Kamiokande \cite{Iked}, MACRO \cite{Ambr}, LVD \cite{Agli2}, AMANDA \cite{Ahre} and SNO \cite{Ahar}. Over the years, our understanding of how massive stars explode and how the neutrino interacts with hot and dense matter has increased by a tremendous degree. At present the scale and sensitivity of the detectors capable of identifying neutrinos from a Galactic supernova have grown considerably so that current generation detectors \cite{Lund,Bell,Eguc} are capable of detecting of order ten thousand neutrinos for a supernova at the Galactic Center.

So a neutrino flux from a next SN will be measured with several detectors that provides unprecedented reliability of obtained information.

The Baksan Underground Scintillation Telescope (BUST) \cite{Alek2} is the multipurpose detector intended for wide range of investigations in cosmic rays and particle physics. One of the current tasks is the search for neutrino bursts. The facility has been uninterruptedly used for this purpose since the middle of 1980. The total observation time of the Galaxy amounts to 90\% of the calendar time. The paper is built as follows. Section 2 is the brief description of the facility. Section 3 is devoted to the neutrino burst detection method and some characteristics of background events. In Section 4, we present the use of two parts of the BUST as two independent detectors (this allows to increase the target mass). The process of generating a warning about a neutrino burst is described in Section 5. Conclusion is presented in Section 6.

\section{The facility}
\begin{figure}
\includegraphics[width=.7\textwidth]{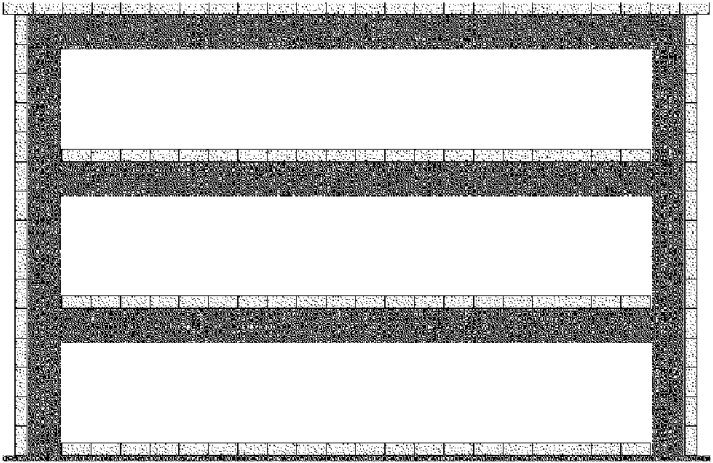}
\caption{The Baksan underground scintillation telescope, side view.}
\label{fig1}
\end{figure}

{\label{one}}
The Baksan Underground Scintillation Telescope is located in the Northern Caucasus (Russia) in the underground
laboratory at the effective depth of $8.5\times 10^4\ g\cdot
cm^{-2}$ (850 m of w.e.) \cite{Alek2}. The facility has dimensions
$17\times 17\times 11$ m$^3$ and consists of four horizontal
scintillation planes and four vertical ones (Fig.~\ref{fig1}).
The upper horizontal plane has an area of 290 $m^2$ and consists of 576 ($24\times
24$) liquid scintillator counters of the standard type, three lower planes have 400 ($20\times 20$) counters each. The distance between neighboring horizontal scintillation layers is 3.6 m. The vertical planes have $15\times 24$ and $15\times 22$ counters. There are another 28 counters which close the slit between the front vertical plane and two side ones.  

The horizontal scintillation planes are located on the floors that consist of an 8-mm-thick iron bottom plate, steel beams (the total iron thickness is 2.5 cm or 20 g cm$^{-2}$), and a 78-cm-thick fill of low-background rock (dunite) (a concrete cap is at the top). The total thickness of one telescope layer
(the scintillator layer plus the floor) is 165 g cm$^{-2}$. The vertical walls of the BUST building are also composed of dunite with iron reinforcement. The charge and atomic weight of the nuclei of BUST material atoms averaged over the volume of one facility layer are $\ove Z$ = 12.8 and $\ove A$ = 26.5, respectively. The radiation unit of length for the telescope material is $t_0$ = 23.5 g cm$^{-2}$.

The standard autonomous counter is an aluminum tank $0.7\times 0.7\times 0.3\ m^3$ in size, filled with an organic $C_nH_{2n+2}$ ($n\simeq 9$) scintillator. The scintillator volume is viewed by one FEU-49 photomultiplier (PM) with a photocathode diameter of 15 cm through a 10-cm-thick organic glass window (the thick window serves to reduce the light collection nonuniformity).

%The average angular resolution of the facility is 2.5$^o$ (it depends on a particle track length), time resolution is 5 ns (i.e. the flight time of a relativistic particle between scintillation planes is measured with an accuracy of 5 ns).

Four signals are taken from each counter. The signal from the PM anode is used to measure the plane trigger time and the energy deposition up to 2.5 GeV (the most probable energy deposition of a muon in a counter is 50 MeV $\equiv $ 1~relativistic particle). The anode signals from the counters of each plane are successively summed in three steps: $\Sigma $25, $\Sigma $100, and $\Sigma $400. In addition to the signals from the entire plane, this also allows the signals from its parts to be used. The current output (the signal from the PM anode through an integrating circuit) is used to adjust and control the PM gain. The signal from the 12th dynode is fed to the input of a discriminator (the so-called pulse channel) with a trigger threshold of 8 and 10 MeV for the horizontal and vertical planes, respectively. The signal from the fifth PM dynode is fed to the input of a logarithmic converter, where it is converted into a pulse whose length is proportional to the logarithm of the signal amplitude \cite {Bak}. The logarithmic channel (LC) allows the energy deposition in an individual counter to be measured in the range 0.5--800 GeV.

In case of cascades initiated by cosmic ray muons, we use the four horisontal scintillation layers (along with overlaps between them) as a 4-row calorimeter \cite {Bak2}. The logarithmic channels allow to measure the longitudinal development and spatial structure of cascades (see Section \ref{two}).

The signal from each plane $\Sigma $400 is fed to linear coders which have the  measurement range of $6 - 80$ MeV and the energy resolution 60 keV. These coders allow us to measure with high accuracy an energy deposition amplitude of single events (see below) which will appear in case of a neutrino burst.

The trigger is an operation of any counter pulse channel of the BUST. The trigger count rate is 17 $s^{-1}$ . When a trigger appears, the entire information on this event is fed to an online computer, in which the events are preprocessed and rendered as a "kadr". The kadr duration is 300 ns (i.e., all the counters that have hit within 300 ns since the trigger moment fall into one kadr). The average kadr processing time ("dead time" of the BUST) is $\simeq $ 1 ms.
 The operation of the individual counters and the data acquisition system is controlled with numerous (about 30) monitor programs, which provide a high reliability of the information from the facility. The GPS signal is used for synchronization with the Universal Time; the synchronization accuracy is 0.2 ms.

\section{The method of neutrino burst detection}
The BUST consists of 3184 standard autonomous counters. The total scintillator mass is 330 t, and the mass enclosed in three lower horizontal layers (1200 standard counters) is 130 tons. The majority of the events recorded with the Baksan telescope from a supernova explosion will be produced in inverse beta decay (IBD) reactions
\begin{equation}
\ove \nu_e + p \to n + e^+ 
\label {1}
\end{equation} 
If the mean antineutrino energy is $E_{\nu _e} = 12 - 15$ MeV \cite{Imsh1,Hill} the path of $e^+$ (produced in reaction (\ref{1})) will be confined, as a rule, in the volume of one counter. In such case the signal from a supernova explosion will appear as a series of events from singly triggered counters (one and only one counter from 3184 operates; below we call a such event "the single event") during the neutrino burst.  The search for a neutrino burst consists in recording of single events cluster within time interval of $\tau $ = 20 s (according to the modern collapse models the burst duration does not exceed 20 s).

The expected number of neutrino interactions detected during an interval of duration $\Delta t$ from the beginning of the collapse can be expressed as: 
\begin{equation}
N^H_{ev} = N_H\int_0^{\Delta t} dt \int_0^{\infty} dE\ F(E,t)\cdot \sigma(E) \eta(E), 
\label {2}
\end{equation}
here $N_H$ is the number of free protons, $F(E,t)$ is the flux of electron antineutrinos, $\sigma(E)$ - the IBD cross section, and $\eta(E)$ is the detection efficiency. The symbol "H" in left side indicates that the hydrogen of scintillator is the target. In calculating (\ref{2}), we used the Fermi-Dirac spectrum for the $\ove \nu_e$ energy spectrum integrated over time (with the antineutrino temperature $k_BT=3.5$ MeV) and the IBD cross section, $\sigma(E)$, from \cite{Strum}.

For an SN at a "standard" distance of 10 kpc, a total energy radiated into neutrinos of $\eps_{tot} = 3\times 10^{53}$ erg, and a target mass of 130 t (the three lower horizontal planes, see Fig. \ref{fig1}), we obtain (we assume the $\ove \nu_e$ flux is equal to $1/6\times \eps_{tot}$ )
\begin{equation}
N^H_{ev} \simeq 38\ \ \ \ (no\ oscillations)  
\label {4}
\end{equation}
Flavor oscillations are unavoidable of course. However, it was recognized
in recent years that the expected neutrino signal depends strongly on the oscillation scenario (see e.g. \cite{Pant,Sawy,Duan,Tamb}).
The oscillation effects depend on many unknown or poorly known factors . These are the self-induced flavor conversions, the matter suppression of self-induced effects, specific flavor conversions at the shock-fronts, stochastic matter flows fluctuations.

In the absence of a quantitatively reliable prediction of the flavor-dependent fluxes and spectra it is difficult to estimate the oscillation impact on $\nu _e$ and $\ove \nu _e$ fluxes arriving to the Earth. Therefore, it is an open question how the estimation (\ref{4}) is changed under the influence of flavor conversions effects.

Background events are i) radioactivity (mainly from cosmogeneous isotopes) and ii) cosmic ray muons if only one counter from 3184 hit. The total count rate from background events (averaged over the period of 2001 -- 2018 years) is $f_1 = 0.0207\ s^{-1}$ in internal planes (three lower horizontal layers) and $\simeq $ 1.5 $s^{-1}$ in external ones. Therefore three lower horizontal layers are used as a target; below, we will refer to this counter array as the D1 detector (the estimation (\ref {4}) has been made for the D1 detector).

\begin{figure}[ht]

   \centering
   \begin{subfigure}[]{0.45\textwidth}
       \includegraphics[width=\textwidth]{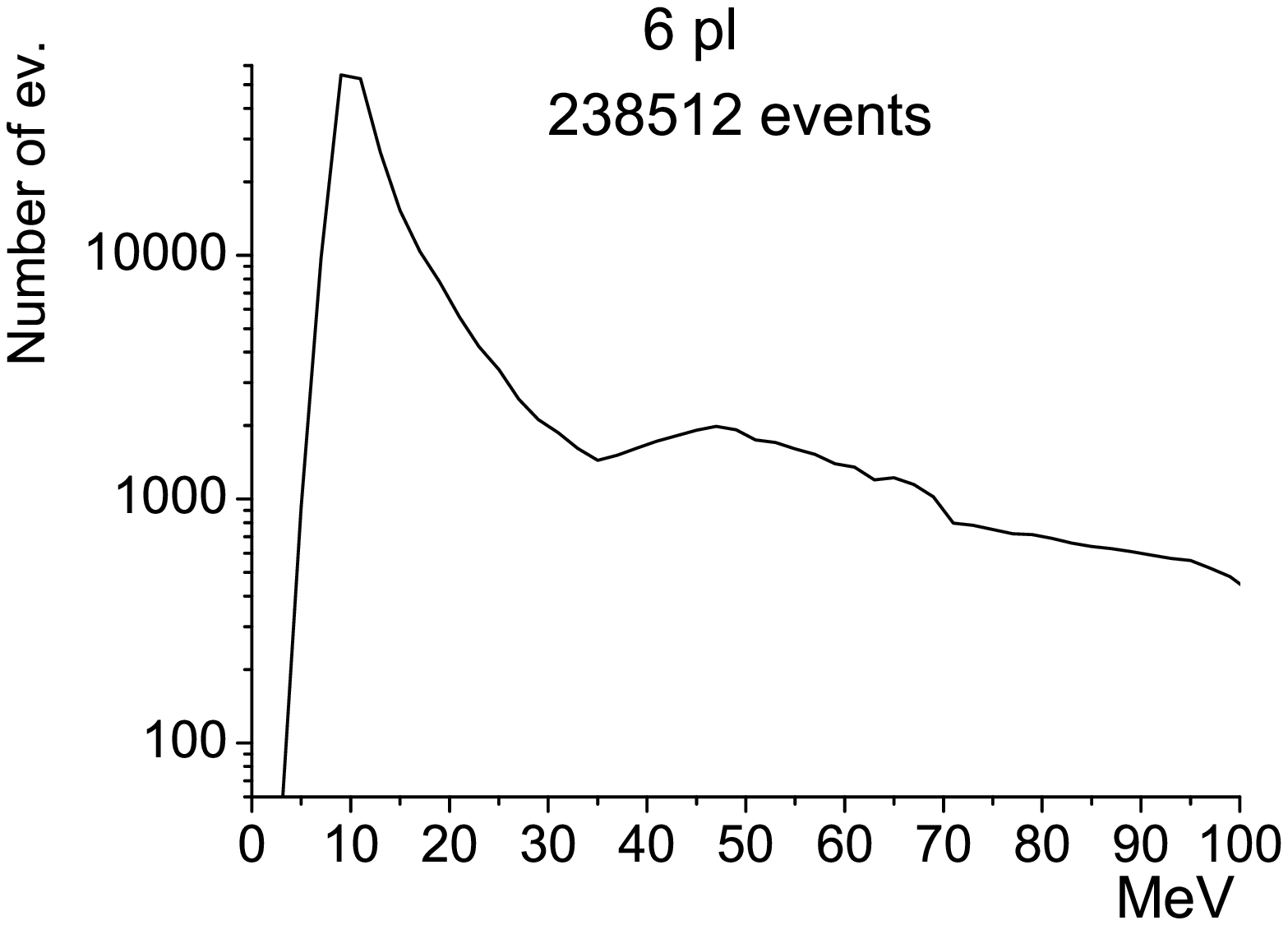}
%       \caption{6 pl}
       \label{fig:3a}
   \end{subfigure}

   \begin{subfigure}[]{0.45\textwidth}
       \includegraphics[width=\textwidth]{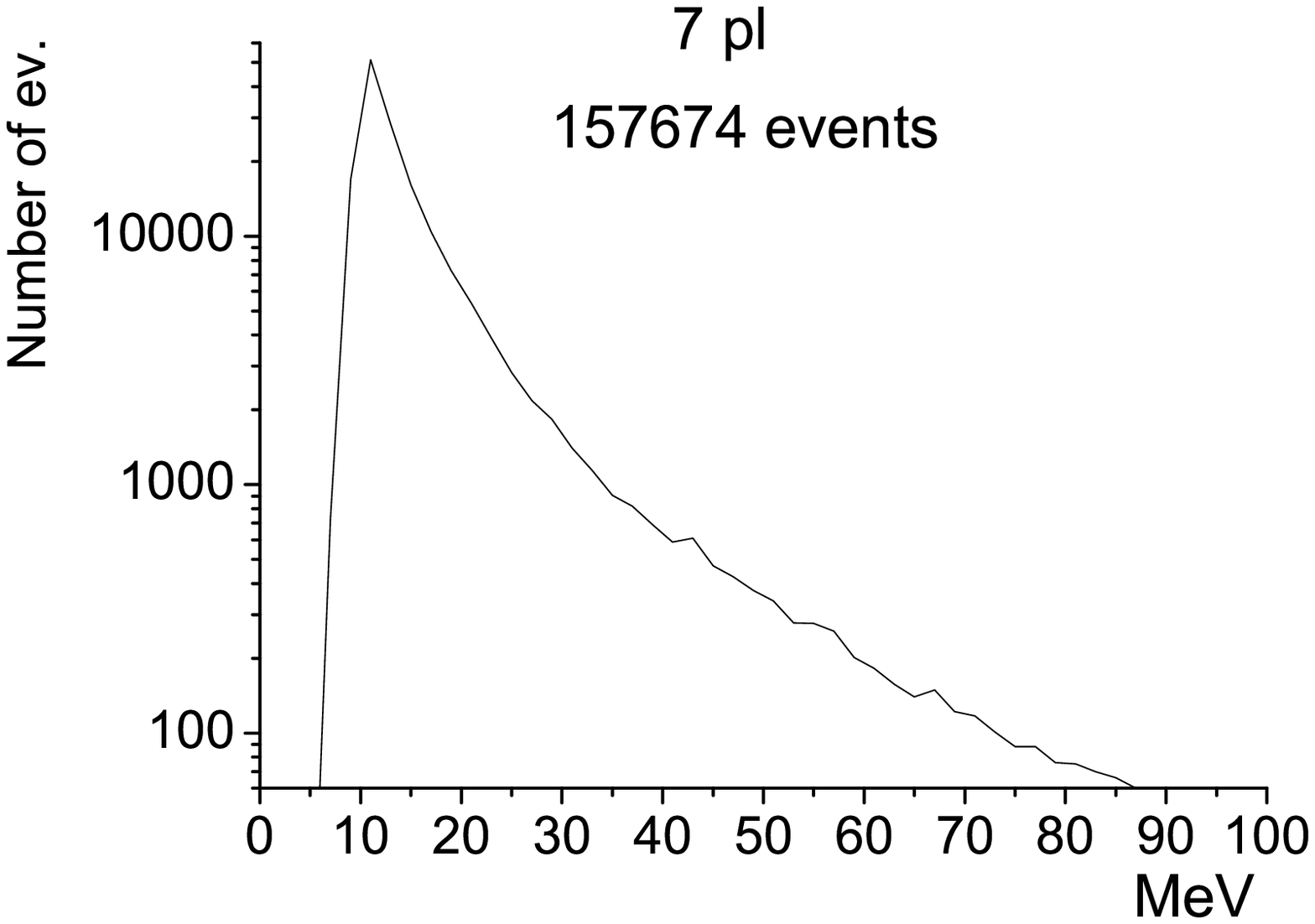}
%       \caption{7 pl}
       \label{fig:3b}
   \end{subfigure}
   ~ %add desired spacing between images, e. g. ~, \quad, \qquad, \hfill etc. 
   %(or a blank line to force the subfigure onto a new line)
   \begin{subfigure}[]{0.45\textwidth}
       \includegraphics[width=\textwidth]{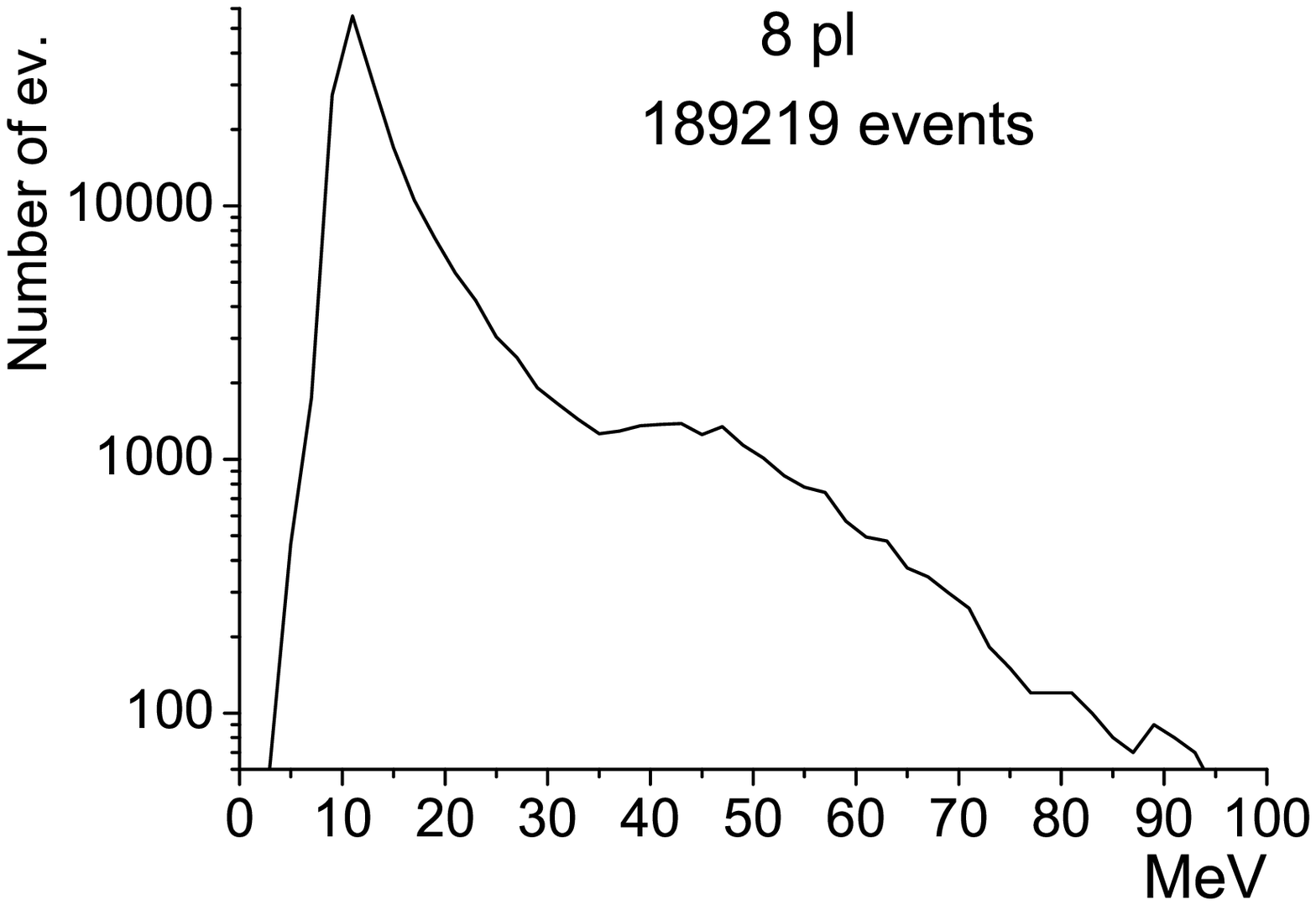}
%       \caption{8 pl}
       \label{fig:3c}
   \end{subfigure}
   \caption{Energy spectra of single events for planes 6, 7 and 8. The exposure time is 322 days. The energy bin width is 2 MeV. The total number of events in the spectrum is indicated on each panel.}
   \label{fig2}

\end{figure}

Figure \ref {fig2} presents the energy spectra of single events (i.e., these are the spectra of background events) for the three lower horizontal scintillation planes; their numbers are 6, 7, and 8 (the numbering from bottom to top). The exposure time is 322 days. The spectra were measured with the linear coders (see Section \ref{one}).

On the sixth and eighth planes a "muon peak" is seen in the region $40-50$ MeV; on the seventh plane this peak is suppressed due to better protection from atmospheric muons.

The peak in the region $10-15$ MeV is related to the decays of cosmogenic isotopes ( $^{12}B,\ ^{12}N,\ ^{8}B,\ ^{8}Li $ etc.), which are produced in inelastic interactions of muons with scintillator carbon and atomic nuclei of the surrounding material.
%\begin{equation}
%^{12}C + \pi^- \rightarrow ^{12}B + \pi^o 
%\label {B}
%\end{equation}
%\begin{equation}
%^{12}C + \pi^+ \rightarrow ^{12}N + \pi^o
%\label {N}
%\end{equation}
 Actually, we observe the combined decay curve from all cosmogenic isotopes that is truncated on the left by the BUST counter trigger threshold (8 MeV); therefore, there is a peak at low energy depositions in the measured spectrum.

We estimated the production rate of unstable isotopes based on the results from \cite{Bell2}. According to this estimate, cosmogenic isotopes create $\simeq $ 50000 single events on each scintillation plane in the exposure time (322 days). The energy deposition from the decays of all isotopes is less than 20 MeV. The remaining single events are created by those muons that pass through the external BUST planes without detection (through the slit between the counters ($\simeq 1$ cm) or touching the counters in such a way that the energy deposition in them is below the 8-MeV threshold) and cause only one counter to be triggered on one of the internal planes.

\begin{figure}[h]
\includegraphics[width=0.9\textwidth]{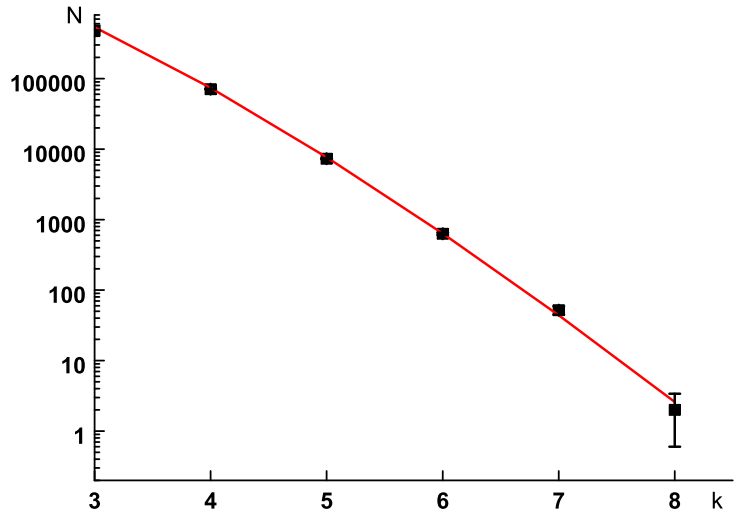}
\caption{The number of clusters with $k$ single events
within time interval of $\tau$ = 20~s. Squares are experimental
data, the curve is the expected number according to the expression (\ref {10})}
\label{fig3}
\end{figure}

Background events can imitate the expected signal (k single events within sliding time interval $\tau$) with a count rate
\begin{equation}
p(k) = f_1\times exp(-f_1\tau)\frac{(f_1\tau)^{k-1}}{(k-1)!} 
\label {10}
\end{equation}
The treatment of experimental data (single events over a period
2001 - 2018 y; $T_{actual}$ = 15.48 years) is shown by squares in Fig. \ref {fig3} in comparison with the expected distribution according to the
expression (\ref {10}) calculated at $f_1 = 0.0207\ s^{-1}$. Note that there is no
normalization in Fig. \ref {fig3}.

It should be explained that the sliding 20-s time interval moves in discrete steps from one single event to the next, so that at least one event is always present in the cluster (at the beginning of the interval). This gives rise to the coefficient
$f_1$ in the expression (\ref {10}). If a new single event falls into the 20-s window when the beginning of the interval passes to the next event, then the number of clusters with a given multiplicity increases by one. If, however, no new event is added and the newly formed cluster has a multiplicity smaller by one than the preceding one, then this cluster is considered to be a fragment of the preceding cluster and is disregarded in the distribution.

This variant of processing guarantees against the loss of a cluster with a greater multiplicity (because some of the events may fall into the neighboring cluster), but, at the same time, some clusters overlap in time, which leads to some deviation from the Poisson distribution.

According to the expression (\ref {10}), background events create  clusters with k = 8 with the rate 0.178 $y^{-1}$. The expected number of such clusters during the time interval T = 15.48 y is  2.75 that we observe (2 events). The formation rate of clusters with k = 9 background events is $9.2\times 10^{-3}\ y^{-1}$, therefore the cluster with multiplicity $k\ge k_{th} = 9$ should be considered as a neutrino burst detection.

\section{Two independent detectors}
\begin{figure}[ht]
\includegraphics[width=0.8\textwidth]{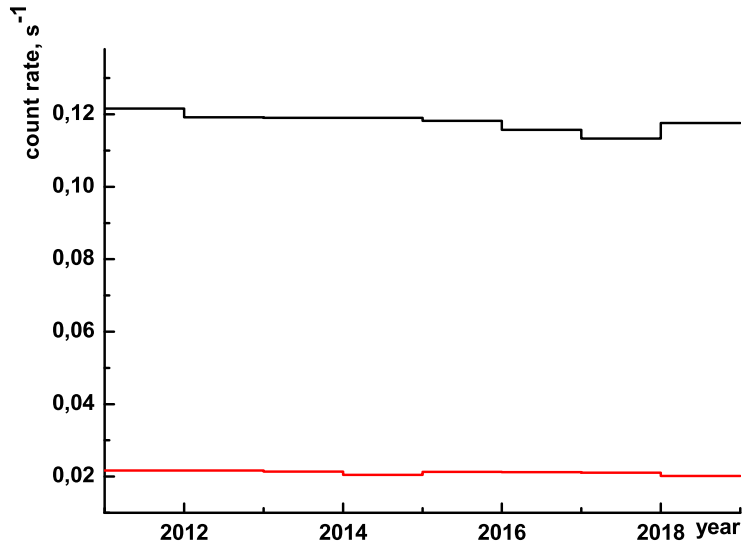}
\vspace{-10mm}
\caption{The count rates of single events in the D1 and the D2 detectors}
\label {fig4}
\end{figure}

{\label{two}}
To increase the number of detected neutrino events and to increase the "sensitivity radius" of the BUST, we use those parts of external scintillator layers that have relatively low count rate of background events. The total number of counters in these parts is 1030, the scintillator mass is 110 tons. We call this array the D2 detector, it has the count rate of single events $f_2$ = 0.12 $s^{-1}$. The count rates of single events in D1 and D2 detectors and the operating stability have been shown in Fig. \ref{fig4}.

The joint use of D1 and D2 detectors allows us to increase the number of detected neutrino events and the detection reliability of a neutrino burst.

We use the following algorithm: in case of a cluster detection with $k1\ge 6$ in   the D1, we check the number of single events, $k2$, in the 10-second time frame in the D2 detector. The start of the frame coincides with the start of the cluster in D1. Mass ratio of D2 and D1 detectors 1030/1200 = 0.858 implies that for the mean value of neutrino events $k1 = 6$ in D1, the mean number of neutrino events in D2 will be $\ove {k2} = 6\times 0.858\times 0.8 = 4.12$ (factor 0.8 takes into account that the frame duration in D2 is 10 seconds instead of 20 seconds in D1). Since the background adds $f_2\times 10$ s = 1.2 events, we obtain finally $\ove {k2}(\ \ove {k1}=6) = 4.12+1.2 = 5.32$.  

According to the exp. (\ref{2}), the expected average number of detected neutrino events in the D2 detector is \ $N^H_{ev}(D2) \simeq 29 $\ (under the same conditions and assumptions as in (\ref{4}) ). So the expected total number of detected neutrino events (in IBD reactions (\ref{1}) ) reads
\begin{equation}
N^H_{ev} = N^H_{ev}(D1) + N^H_{ev}(D2) \simeq 67\ \ \ \ (no\ oscillations) 
\label {5}
\end{equation}
 
The D1 and D2 detectors are independent, therefore the imitation probability of clusters with multiplicities k1 in D1 and k2 in D2 by background events is the product of appropriate probabilities
\begin{equation}
P(k1,k2) = P1(k1)\times P2(k2)
\label {9}
\end{equation}
and we obtain $P(6,5) = 0.23\ y^{-1}$, $P(6,6) = 0.045\ y^{-1}$ (note that $P1$ is determined according to the expression (\ref{10}) and $P2$ is the Poisson distribution for the 10-second time frame).

Therefore the events with $k1\ge 6$, $k2\ge 6$ should be considered as
candidates for a neutrino burst detection (since mean values of k1 and k2 are significantly exceeded in two independent detectors simultaneously and the imitation probability of such events by background is very small).

Notice that in case of a real neutrino burst, the remain counters (which do not belong to D1 and D2) can be used as the third independent detector -- D3 with the mass of 100 ton. Since the count rate of background events in the D3 is $f_3 \simeq$~1.4~$s^{-1}$, the D3 can only be used if the multiplicity of clusters is large enough: $k1\ge$ 10, $k2\ge$ 10 (in the time frame of 20 seconds). Then the multiplicity of the cluster in the D3 should be $k3\ge$ 30. However, we have to eliminate background events in the D3. The energy spectra of clusters in the D1 and D2 can be used for that.

Below, we present an example of a cluster for which combined use of D1 and D2 detectors has the significant advantage. On the 31st of October, 2017 the cluster with the multiplicity $k=9$ was recorded in the D1. UTC of the first single event of the cluster was 17 hours 51 min 28.7602 s. We have checked the cluster prehistory (it is the standard procedure). The nuclear cascade was 7 ms before the first event of the cluster. In case of cascade events, we can use the horisontal layers of the BUST as a 4-row calorimeter (see e.g. \cite{Bak2}). The cascade had the energy $E = 12.3$ TeV, the energy depositions in scintillator layers (from top to bottom, see Figure~\ref{fig1}) were 1.2 GeV, 265 GeV, 1090 GeV, and 355 GeV. 

\begin{table}[ht]
\begin{center}
\caption{Parameters of the cluster with $k = 9$ in the D1 detector. $\Delta t$ is the delay time of the event relative to the cascade; see the text.}
\begin{tabular}{|c|c|c|c|}
\hline $n_{ev}$ & $\Delta t$, s & counter & $\eps$, MeV \\\hline
1 & 0.007 & 7:9-14 & 12.1 \\\hline
2 & 0.015 & 6:15-16 & 11.7 \\\hline
3 & 0.023 & 6:17-16 & 12.1 \\\hline
4 & 0.080 & 6:15-12 &  9.8 \\\hline
5 & 0.101 & 6:18-14 & 13.4 \\\hline
6 & 0.131 & 6:20-13 & 10.7 \\\hline
7 & 0.775 & 7:13-16 & 12.0 \\\hline
8 & 1.341 & 7:12-16 & 9.0  \\\hline
9 & 19.529 & 6:8-13 & 38.7 \\\hline
\end{tabular}
\label{tab1}
\end{center}
\end{table}
The cluster parameters are shown in the table \ref{tab1}: $n_{ev}$ is the event number in the cluster, $\Delta t = t_{ev} - t_c$ is the delay time of the event relative to the cascade, the column "counter" is counter coordinates (for example 7:9-14 means Plane 7, Row 9, Column 14), $\eps$ is the energy deposition in the counter.

As can be seen from the table, the first 8 events occurred during 1.35 s after the cascade and were found in the planes with greatest energy depositions (1090 GeV in Plane 7 and 355 GeV in Plane 6). Therefore we assume that all 8 events are the decays of cosmogenic isotopes produced in the cascade. At the same time, 3 single events occurred in the D2 detector (with $\Delta t$ = 1.045 s, 5.761s, 8.807 s). The absence of single events in Plane 8 and a small number of events in the D2 detector confirm this conclusion. 

Therefore one should not consider this cluster as a candidate for a neutrino burst (since we have established its origin) and it is not included into the
distribution in Figure 3.

It should be mentioned that the cluster with $k=9$ has been recorded only once over
all the observation time ($\simeq $ 38 years).

We have checked the prehistory of all clusters with multiplicities $k\ge 6$. No clusters have been found in which $\ge 5$ single events coincided with a previous cascade within the frame of 2 seconds.

\section{The SN neutrino burst warning}{\label{five}}
The SN monitor system scheme described in this Section came into service in June of 2016. Before that time, the monitor system had been running as one of the offline processes.

The BUST data collected by the data acquisition system are sent to the on-line computer operative memory. Every 15 minutes (0, 15, 30 and 45 minutes of every hour) this information is written down in the file which number is in a one-to-one correspondence with the calendar time. We call it a RUN-file. In 20 seconds the RUN-file is sent to the off-line computer where in 4 minutes the off line data processing begins (among others things, the search for neutrino burst according to the algorithm described in Section \ref{two}). This process takes about 1 minute, thus we obtain the result within 20 minutes (if the neutrino burst has occured in the beginning of the 15-minutes time interval). When the process finds the event $k1\ge 6$, $k2\ge 6$ (see Section \ref{two}), it generates a SN warning which initiates phone-callings and emails sent to experts in the BUST collaboration. The experts make a decision to make a world-wide announcement or not within one hour. We plan also to connect our facility to the SuperNova Early Warning System (SNEWS) \cite{Anto}.

It should be noted that in the case of a very close SN, for example at the distance of d = 0.2 kpc, the total number of events from IBD reactions, according to the estimate (\ref{5}), will be $\simeq $~250,000. In the first seconds (after a core bounce), we should expect $\simeq$ 30,000 events per second. Against this count rate, the count rate of muon events (17 s$^{-1}$) is negligible. So all events recorded by the BUST (with all 3184 counters, the scintillator mass is 330 ton) during this time period will be neutrino events. The kadr duration of the BUST is 300 ns, the kadr processing time is $\simeq $ 1 ms, therefore we will record $\simeq $ 1000 events per second, with the overwhelming majority of events being kadrs with one counter. The fraction of kadrs in which two counters hit (i.e. two neutrino events fell in the time frame of 300 ns) is less than 0.5\%. Thus, in the case of a very close SN, some part of the events (which depends on the distance to the SN) will be lost.

\section{Conclusion}
The Baksan Underground Scintillation Telescope operates under the program
of search for neutrino bursts since June 30, 1980. As the target, we use two parts of the BUST (the D1 and D2 detectors) with the total mass of 240 tons. The estimation (\ref{5}) allows us to expect $\simeq $ 10 neutrino interactions from a most distant SN ($\simeq $ 25 kpc) of our Galaxy. In the opposite case, of a very close SN, some part of the events (which depends on the distance to the SN) will be lost (see Section~\ref{five}).

Background events are 1) decays of cosmogeneous isotopes (which are produced in inelastic interaction of muons with the scintillator carbon and nuclei of surrounding matter) and 2) cosmic ray muons if only one counter from 3184 hit.

Over the period of June 30, 1980 to December 31, 2018, the actual observation time was 33.02 years. This is the longest observation time of our Galaxy with neutrino at the same facility. No candidate for the core collapse has been detected during the observation period. This leads to an upper bound of the mean frequency of star  gravitational collapses in the Galaxy 
\begin{equation}
f_{col} < 0.070\ y^{-1}
\label {11}
\end{equation}
at 90\% CL. Recent estimations of the Galactic core-collapse SN rate give roughly the value $\simeq 2-5$ events per century (see e.g. \cite{Adam}).

The work has carried out at a unique scientific facility the Baksan Underground Scintillation Telescope (Common-Use Center Baksan Neutrino Observatory INR RAS) and was supported by the Program for Fundamental Scientific Research of RAS Presidium  "Physics of hadrons, leptons, Higgs bosons and dark matter particles".

\end{document}